\documentstyle[12pt]{article}

\title{Sonoluminescence as a physical vacuum excitation}
\author{Yu.P.Stepanovsky and G.G.Sergeeva}
\date{}
\begin{document}
\maketitle
\thispagestyle{empty}

\begin{abstract}
It is discussed a physical vacuum excitation as a mechanism of
sonoluminescence of gas bubble. This Schwinger's theory
was based on the assumption that the sudden change of the collapse rate of
bubble in the water leads to the jump of the dielectric constant of gas.  It
is shown that the dependence of the dielectric constant on the gas density
really leads to the jump of the dielectric constant at the shock-wave
propagation in a collapsing gas bubble.

\end{abstract}

  Sonoluminescence is a transformation of sound into the light.
The liquid (it can be the distillate water) is incurred to the
acoustic waves, and photons are radiated at that.
The radiated light energy is comparable with introducing
sound energy. The sound energy density is of the order $10^{-11}$ eV/atom,
and as the energy of radiation photons is of the order 10 eV, one can say
that the density of the energy is increases in $10^{12}$ times.  Sonoluminescence is
  well-known and studied for 60 years.  The cavitation nature of the
radiation is determined~\cite{Lev}. The dependence of sonoluminescence
intensity from chemical structure of liquid and dissolved gas in this liquid,
their thermal conductivity coefficients, from temperature, from sound
frequency and amplitude is studied.  In spite of, that after stable
luminescence observation of single gas bubble~\cite{Gai,Bar}, the
sonoluminescence attracted attention of many investigators [4-12], the
conclusive explanation of physical mechanism of this effect is not find out
now.

  The sonoluminescence in Schwinger's works is considered as a
manifestation of nonstationary Casimir effect~\cite{Sch}, which is a result
of the change of vacuum properties under different external influences.
At the studying of sonoluminescence the gas bubbles (cavities in the
water with radius $r$, which are filled up by the gas) are putting to
compression and expansion in the response to the positive and negative
changes of pressure under the acoustic field influence. Schwinger's
theory is based on the assumption
that the sudden change of the collapse rate of bubble in the water
leads to the jump of its dielectric constant. This is accompanied by the
excitation of electromagnetic vacuum and by the photons radiation.
In~\cite{Sch}  the probability of the photons creation is calculated
and the numeral accounts which are qualitatively explain the part
of experiment datas, are doing. The task of this paper is to lead
some reasons for the sustention of Schwinger's idea and to show that the
dielectric constant of gas is really jump function of the time at
nonadiabatic stage of the bubble collapse.

  Well-known Schwinger's   expression for the total
number of photons created in the volume V~\cite{Sch}:
\begin{equation}\label{1}
N=\int\limits_{}^{}\frac{(\sqrt[]{\varepsilon}-1)^2}{4~\sqrt[]{\varepsilon}}
\frac{d^3k}{(2\pi)^3}V.
\end{equation}
can be obtained as solution of an equation, which is described the
scalar field of photons~\cite{Step}. This fact is evidenced
about the validity of Schwinger's idea.
 For removing the integration divergency in the formula (1) we
must to make cut-off wave-number $k_{max}$, which corresponds to the
minimum wave length $\lambda_{min}<r_m$ ( $r_m$ is a bubble radius at the
moment of photons radiation). Disregarding the dispersion, Schwinger obtained
  the following expression for the energy $E$ emitted in unite volume:
\begin{equation}\label{2}
\frac{E}{V}=\frac{(\sqrt[]{\varepsilon}-1)^2}{4~\sqrt[]{\varepsilon}}
\int\limits_{}^{k_{max}}\hbar\omega\frac{d^3k}{(2\pi)^3}=
2{\pi}^2\frac{(\sqrt[]{\varepsilon}-1)^2}{4~\sqrt[]{\varepsilon}}
\frac{\hbar c}{\lambda^4_{min}}.
\end{equation}

The expression (2) is in qualitative accordance with the result of
Ref~\cite{Milt}, which was obtained at the divergency  removal by taking into
account the contribution of the surface energy. This is
the serious substantiation of the validity for conducted by Shwinger
removal of the divergency.

  The value $\varepsilon\approx1$ for the many gases in normal conditions at
adiabatic compression. From formula (2) one can see
that the number of photons and the emitting energy are little
as $(\sqrt[]{\varepsilon}-1)^2$. But the value of gas temperature within
the bubble (~$10^5$ K) which observed in the experiment, compelled to
remember the Jarman's idea~\cite{Jar}  about excitation of gathering shock
wave at the nonadiabatic stage of bubble collapse~\cite{Green,Wu}.  The
numerical solution of Rayleigh-Plesset equation of a bubble surface, of van
der Waals equations for the gas inside bubble and of tasks about generation
and  motion of the shock wave was carried out in a
work~\cite{Wu}.  At a first stage of a bubble growth it was used an adiabatic
approximation:  by the time $\sim16,65 \mu s$, the bubble radius increased
   from the initial value $r_0=4,5 \mu m$ to the maximum value $\sim 37,09
\mu m$ and the gas density decreased to the value $0,0023~kg~m^{-3}$. At the
second stage of a bubble collapse calculations were done in a nonadiabatic
approximation, which leads to the Guderley's decision for an ideal
gas~\cite{Gud}. As it was shown in [10], at nonadiabatic stage of collapse at
the focusing shock wave in the centre of bubble the density of nonideal gas
increases in $10^{5}$ times by the time $\sim 3,84 \mu s$, and achieving the
maximum value $\rho \sim \rho_m \sim 794 kg~m^{-3}$. This conducts by the
change of the rate of bubble collapse to $\sim 2\cdot 10^{4}m~ s^{-1}$.
Further it will be shown that it leads to the jump of the dielectric constant
of gas.

  The value $\varepsilon$ for gas is determined by Clausis-Mossotti formula
~\cite{Fr}
\begin{equation}\label{3}
\varepsilon-1=\frac{4\pi p\rho}{w(1-\frac{4\pi p\rho}{3w})},
\end{equation}
where $p$ is the molar polarizability of gas, $\rho$ is the gas density
and $w$ is his molecular weight. The liquid which surrounds the gas bubble
(in the experiment this is the water with small addition of glycerin)
leads to the additional polarization and to the molar polarizability of gas
dependence on the dielectric constant of liquid $\varepsilon_1$:
\begin{equation}\label{4}
p=\frac{p_0}{3(1+\frac{\varepsilon}{2\varepsilon_1})}.
\end{equation}
where $p_0$ is the molar polarizability of gas in normal conditions.
Even taking into account that
$\rho_m\sim \rho$ at focusing shock wave,
the ratio $\varepsilon/\varepsilon_1 << 1$, so
\begin{equation}\label{5}
\varepsilon(t)-1\approx\frac{4\pi p_0\rho(t)}{3w}.
\end{equation}
Such dependence of $\varepsilon(t)$
on $\rho(t)$ leads to the  jump of the dielectric constant
at the  jump of the gas density as shock wave focuses at the centre
of bubble. As we can see from (2) this jump accompanies by
 the vacuum excitation and to the photons
radiation, the intensity of which depends from
the characteristic time of the  jump of the gas density.

\end{document}